\documentclass[%
 reprint,
 amsmath,amssymb,
 aps,
]{revtex4-2}

\usepackage{graphicx}
\usepackage{dcolumn}
\usepackage{bm}
\usepackage{hyperref}

\begin{document}

\preprint{APS/123-QED}

\title{Resolving Mutually Coherent Bright Point-Sources}

\author{Ilya Karuseichyk}
\email{ilya.karuseichyk@lkb.upmc.fr}
\author{Giacomo Sorelli}
\author{Manuel Gessner}
\author{Mattia Walschaers}
\author{Nicolas Treps}
\affiliation{Laboratoire Kastler Brossel, Sorbonne Universit\'{e}, CNRS, ENS-Universit\'{e} PSL, Coll\`{e}ge de France, 4 place Jussieu, F-75252 Paris, France
}%


\date{\today}

\begin{abstract}
 We analyze the problem of resolving two point-sources in the case of mutually coherent sources with arbitrary quantum statistics, mutual phase, relative and absolute intensity. We use a sensitivity measure based on the method of moments and compare direct imaging with spatial mode demultiplexing, analytically proving advantage of the latter.
 We show that the sensitivity of spatial mode demultiplexing saturates the quantum Fisher information, for all known cases, even for non-Gaussian states of the sources. 

\end{abstract}

\maketitle


\section{Introduction}
The problem of resolving two point sources has been intensively studied as model problem for optical system resolution characterisation for more then a century. For a long time only visual observation of the light coming from the sources was available, thus resolving criteria were based on visual characteristics of intensity distribution \cite{Rayleigh1879,sparrow1916spectroscopic}. Visual criteria, such as Rayleigh criterion, were aimed to distinguish between images of two sources and one, but not to describe the problem of the separation estimation. 

Later it became possible to carry out an analysis of the whole intensity distribution in the image plane, that allows to resolve sources beyond visual limits \cite{helstrom1967image}. The efficiency of this approach can be estimated on the basis of the Fisher information (FI) 
and the Cram\'er-Rao bound. Measuring intensity distribution is called direct imaging (DI). This technique proved to often lead to vanishing FI in the limit of small separations, this feature is known as "Rayleigh's curse". At the same time analysis of the Quantum Fisher information (QFI), that provides the ultimate limit for resolution regardless of the measurement, shows that in most cases one can increase resolution and avoid "Rayleigh's curse" by choosing more optimal measurement than DI \cite{TsangPRX}. 

A fine example of such measurement is spatial mode demultiplexing (SPADE). SPADE proposes to decompose the field in the image plane into spatial modes and measure their intensities. The resolution of SPADE was widely studied for uncorrelated thermal sources. The FI of this measurement, calculated in the limit of small photon numbers, was proved to saturate the ultimate bound set by QFI \cite{TsangPRL,pirandola}. Recently another measure of resolution was proposed --- a sensitivity measure based on method of moments \cite{PRLGessner2019}. This measure allowed to analyse problem of resolving bright uncorrelated thermal sources and proved SPADE to saturate the ultimate limits without additional assumption about sources intensities~\cite{sorelli2021PRL,sorelli2021PRA}. 

At the same time an important discussion about the role of sources' mutual coherence took place \cite{Larson:18,tsang2019comment,Larson19reply,Hradil19QFI}. 
Two different approaches to describe partial coherence of low brightness sources, departing in their account of losses, have been introduced, leading to different results. Recently it became clear that, for the case of unknown sources' brightness, a loss-less model \cite{Larson:18} can be applied. 
On the other hand, if the sources' brightness is assumed to be known, one needs to rigorously account for losses on the finite aperture of the optical system to benefit from  dependence of the total registered intensity on estimated separation \cite{tsang2019comment,kurdzialek2021sources}. 

Despite these intensive studies of the role of the mutual coherence of the sources there are still gaps in knowledge in this area. This is especially true for bright sources, since most of the cited research is based on the assumption of low intensity of the detected signal. Wherein the study of the QFI of partially correlated thermal light shows a significant influence of the sources' brightness on resolution limit \cite{pirandola}. Thus for practical application of SPADE it is important to develop its description for the case of bright sources and to be able to estimate the sensitivity of specific measurements. 

In this study we focus on the fundamentally and practically important case of a pair of perfectly mutually coherent sources. We consider the most general quantum state of emitted light, and make no assumption about sources' absolute and relative brightness and relative phase.
Using the method of moments, we analyse the sensitivity of DI and SPADE for sources' separation estimation and analytically show an advantage of the latter approach. Considering a multiparameter estimation approach we analytically find the separation estimation sensitivity with and without prior knowledge about the brightness of the sources. We show that anti-bunching of the sources' radiation leads to an increase of the sensitivity, but at the same time ignorance of sources brightness wipes out any possible profit from non-classical statistics of the sources. 

Finally, we show that the calculated sensitivity saturates the QFI for the cases where it is known. These include non-Gaussian entangled states of the sources, for which the QFI of separation estimation is maximal.

Our approach can be easily generalised to other parameter estimation problems with mutually coherent probes or single-mode probe, like coherent imaging \cite{ferraro2011coherent} or even distributed quantum sensing \cite{guo2020distributed}. Moreover, moment-based estimation gives a practical way to estimate a bound on the sensitivity of specific measurements which can be saturated by a parameter retrieval method that does not require complicated data processing.

\section{Optical scheme}
\begin{figure}[ht]
  \includegraphics[width=0.95 \linewidth]{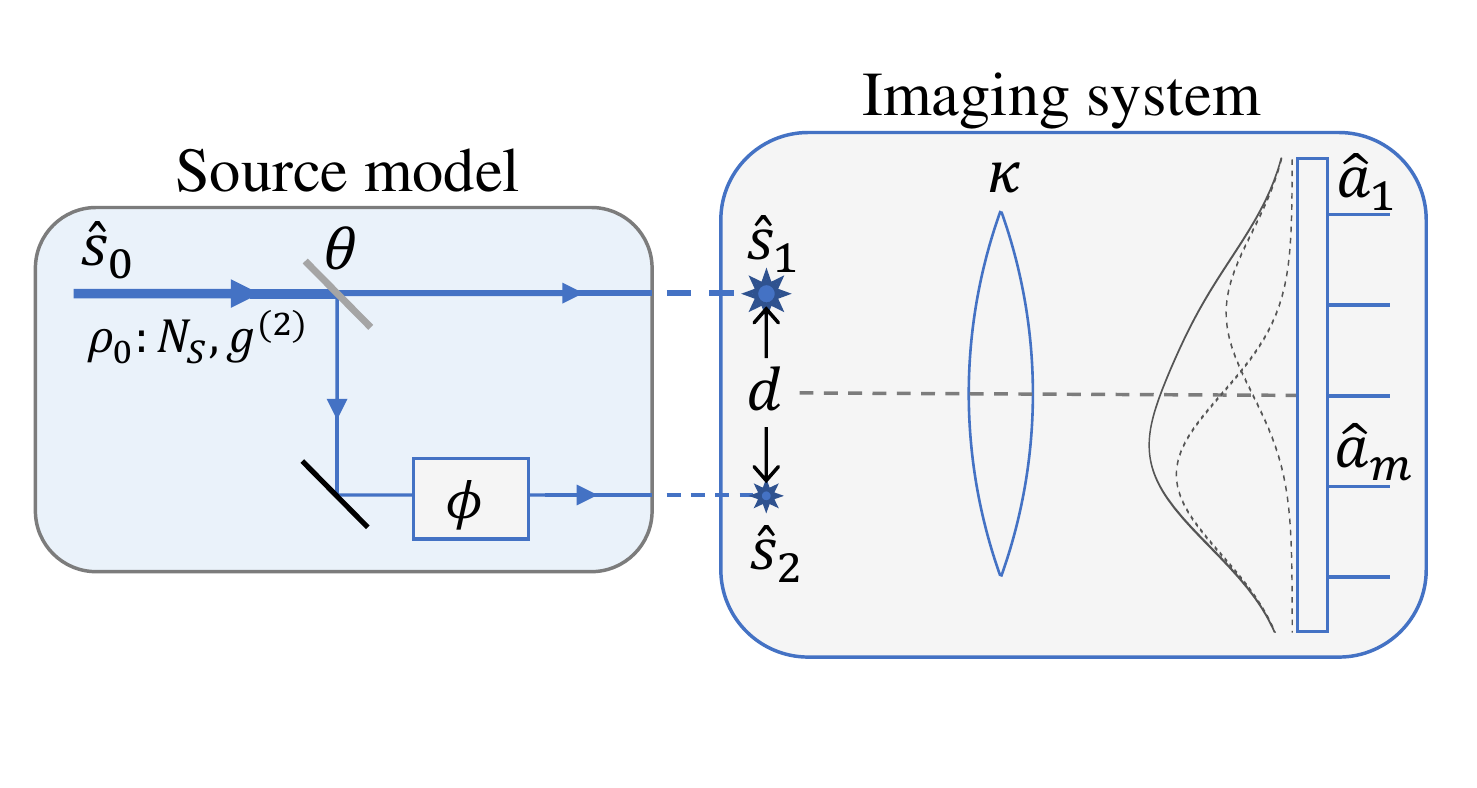}
 \caption{\label{fig:scheme} On the left: principal scheme for generating general two-mode coherent  state based on beamsplitter with transmissivity $T=\cos^2 \theta$ and phase shifting element $\phi$. On the right: optical scheme for sources' separation estimation, were $\kappa$ is the transmissivity and photon counting is performed in the measurement modes $f_m (\vec r)$ with corresponding field operators~$\hat a_m$.}
 \end{figure}

We consider a traditional optical scheme for parameter estimation problem, where two point-like sources emit light, that passes through diffraction-limited imaging system (see the right part of fig.\ref{fig:scheme}). Then sources' parameters, such as separation $d$, are estimated from measuring the diffracted light.

\subsection{Emitted field}
Point-sources emit light in the orthogonal modes with field operators $\hat s_{1,2}$. We only consider mutually coherent sources, meaning that the absolute value of the first order coherence degree  of emitted light is $| g^{(1)}|=1.$ One can always find single mode (called principal mode) that fully describe this field configuration. Thus the most general mutually coherent state of the  modes $\hat s_{1,2}$ can be considered as the result of splitting some mode $\hat s_0$ on an asymmetric beamsplitter with transmissivity $T=\cos^2 \theta$ and adding phase $\phi$ to one of the output modes $\hat s_{1,2}$ (see the left part of fig.\ref{fig:scheme}).  The first order coherency matrix of the modes $\hat s_{1,2}$ reads 
\begin{equation}
    \label{eqn:cohS} 
    \langle \hat s^\dagger_j \hat s_k \rangle=
    \begin{pmatrix}
    N_S \cos^2 \theta & \frac{N_S}{2} \sin 2\theta~e^{i \phi} \\
     \frac{N_S}{2} \sin 2\theta~e^{-i \phi} & N_S \sin^2 \theta
        \end{pmatrix},
\end{equation}
where $N_S=\operatorname{Tr} (\hat \rho_0 \hat s_0^\dagger \hat s_0 )  $ -- average number of emitted photons and parameter $\theta$ is responsible for sources' intensity asymmetry. The ratio between the sources' brightness $\langle \hat s_2 ^\dagger \hat s_2 \rangle/\langle \hat s_1 ^\dagger \hat s_1 \rangle=\tan^2\theta$. For concreteness we assume parameter $\theta$ to be within a range $0 \le \theta \le \pi/4$, where $\theta=\pi/4$ corresponds to equally bright sources, and $\theta=0$ to all light in mode $\hat s_1$.
Parameter $\gamma=\langle \hat s_1^\dagger \hat s_2 \rangle/\sqrt{\langle \hat s_1 ^\dagger \hat s_1 \rangle \langle \hat s_2 ^\dagger \hat s_2 \rangle}=e^{i\phi}$ is often referred to as the degree of mutual coherence between modes, and in our case it has modulus equal to one.  

Note, that no assumptions about state $\hat \rho_0$ of the mode $\hat s_0$ was made so far. If it has Poisson photon-number statistics, the state of the modes $\hat s_{1,2}$ is separable; for thermal statistics of $\hat \rho_0$ the modes $\hat s_{1,2}$ are classically correlated; for Fock state of $\hat s_0$ the output modes are entangled.

 For further description of the parameter estimation in terms of the method of moments we will  only need the two first moments for the state $\hat \rho_0$: average photon number $N_S$ and its variance $\Delta N_S^2$. 

\subsection{Field detection}

The emitted light goes through an imaging system with finite aperture that has transmissivity $\kappa$ and  point spread function (PSF) $u_0(\vec r)$. We use paraxial approximation where loss factor does not depend on position of the source.  In  the image plane the light is detected either directly (DI) or via SPADE. Both cases can be described as measurement over some field modes $f_m( \vec r)$ with corresponding operators $\hat a_m$, where for DI $f_m( \vec r)$ are localized pixel modes and for SPADE $f_m( \vec r)$ are more general non-localized modes, for example Hermite-Gauss. Then the parameters of interest are estimated from measured numbers of photons in detection modes $N_m=\langle \hat a_m^\dagger \hat a_m \rangle$. Particularly, we analyse the estimation of the sources' separation $d$ with and without prior knowledge of the number of emitted photons $N_S$. 

Passing through the lossy optical system can be described as mixing of field modes 
with 
vacuum modes \cite{pirandola}. Taking into account also vacuum mode from beamsplitter $\theta$ the field operators of the measurement modes can be represented as $\hat a_m = A_m \hat s_0 + \hat v_m$, where $A_m$ is a complex coefficients and $\hat v_m$ - non-orthogonal combinations of vacuum modes' field operators 
. Thus in measurement modes any normally ordered average  has the following property:
\begin{equation}
\label{eqn:property}
    \langle : F (\hat a^\dagger_m, \hat a_n): \rangle = \langle : F (A_m^* \hat s_0^\dagger ,  A_n \hat s_0): \rangle ,
\end{equation}
where $:\hat X:$ denotes normal ordering, $F$ is an arbitrary analytical function of the field operators. Then average number of detected photons in the $m$-th measurement mode reads
 \begin{equation}
 \label{eqn:Nm}
     N_m=\langle \hat a_m^\dagger \hat a_m \rangle=|A_m|^2 N_S.
 \end{equation}

Specifically for the scheme on fig.\ref{fig:scheme} the coefficients
\begin{equation}
\label{eqn:Am}
A_m=\sqrt \kappa \int d \vec r~f_m(\vec r) \big[u_0(\vec r - \vec r_1) \cos \theta  + u_0 (\vec r- \vec r_2) e^{i \phi} \sin \theta \big]       
\end{equation}
 depends on the sources positions $\vec r_{1,2}$, their intensity asymmetry (that depends on $\theta$) and the mutual phase $\phi$, optical system transmissivity $\kappa$ and PSF $u_0(\vec r)$ and measurement modes shape $f_m(\vec r)$. 
 
\section{Moment based sensitivity}
To estimate resolution of the considered optical scheme we calculate the moment-based sensitivity (further we call it just sensitivity). Within the framework of the method of moments it was shown that estimation a set of variables $\{q_\alpha\}$ from mean values $\{X_m\}$ of observables $\{\hat X_m\}$  corresponds to sensitivity matrix
 \cite{PRLGessner2019}
\begin{equation}
     \label{eqn:sensitivity}
     M_{\alpha \beta} = \sum_{m,n} (\Gamma^{-1})_{m n} \frac{\partial X_m} {\partial q_\alpha} \frac{\partial X_n} {\partial q_\beta},
 \end{equation}
where $\Gamma_{m n}=\langle \hat X_m \hat X_n \rangle-X_m X_n$ is the covariance matrix of the observables. Inverted sensitivity matrix sets a lower bound for estimators' covariances
\begin{equation}
    \operatorname{cov}[q_\alpha,q_\beta] \ge \frac{1}{\mu} ( M^{-1})_{\alpha \beta},
\end{equation}
where $\mu$ is the number of measurement repetitions.
Generally estimators that includes all the moments of observables can have lower variance, i.e. the following chain of inequalities holds
\begin{equation}
    \label{eqn:inequality}
    \mathbf{M} (\{q_\alpha\}, \{\hat X_m\}) \le  \mathcal{F} ( \{q_\alpha \}, \{\hat X_m \}) \le  \mathcal{F}_Q (\{q_\alpha\}),
\end{equation}
where sensitivity matrix $ \mathbf{M} (\{q_\alpha\}, \{\hat X_m\})$ is given by eqn. \eqref{eqn:sensitivity}, $\mathcal{F} ( \{q_\alpha\}, \{\hat X_m \})$ is Fisher information (FI) matrix, $\mathcal{F}_Q(\{q_\alpha \})$ is quantum Fisher information (QFI) matrix and matrix inequality $\mathbf{A} \le \mathbf{B}$ means that $\vec a^T \mathbf{A} \vec a \le \vec a^T \mathbf{B} \vec a$ for any given column vector $\vec a$. Thus the sensitivity can be considered as a lower bound for the FI. It was shown, that the sensitivity of photon counting in Hermite-Gauss (HG) modes for estimation of separation between equally bright uncorrelated thermal sources saturates the QFI~\cite{sorelli2021PRA}. 

Sensitivity matrix also gives a set of linear combinations of observables $\hat X_m$, that contains sufficient information for estimation of the parameters $\{q_\alpha\}$ with precision set by \eqref{eqn:inequality}. These combinations allow to avoid complicated numerical method of parameters inferring such as maximum likelihood estimation.

Using expression \eqref{eqn:sensitivity} one can calculate the sensitivity of photon counting in spatial modes $ f_m (\vec r)$, i.e. use  $\hat X_m=\hat a^\dagger_m \hat a_m$ as an observables
. Thus
using property \eqref{eqn:property} and equation \eqref{eqn:Nm} one can find the elements of the photon number covariance matrix 
 \begin{equation}
 \label{eqn:gamma}
  \Gamma_{m n}=\delta_{mn} N_m + h N_m N_n,
\end{equation}
where $h=g^{(2)}-1=(\Delta N_S^2 - N_S)/N_S^2$. Matrix \eqref{eqn:gamma} can be analytically inverted with Sherman-Morrison formula \cite{sherman1950adjustment} obtaining 
 \begin{equation}
 \label{eqn:gamma_Inv}
  (\Gamma^{-1})_{m n}=\delta_{mn} N_m^{-1} - \frac{ h}{1+h N_D},
\end{equation}
where $N_D=\sum_m N_m$ is the 
total number of detected photons. Then the sensitivity matrix \eqref{eqn:sensitivity} 
can be expressed as
\begin{equation}
    \label{eqn:Mmatrix}
    M_{\alpha \beta}=N_D \sum_m \frac{1}{\varepsilon _m} \frac{\partial \varepsilon_m}{\partial q_\alpha} \frac{\partial \varepsilon_m}{\partial q_\beta} + \frac{1}{\Delta N_D^2}  \frac{\partial N_D}{\partial q_\alpha} \frac{\partial N_D}{\partial q_\beta} ,
\end{equation}
where $\varepsilon_m=N_m/N_D$ is the relative photon number and
\begin{equation}
\label{eqn:N_D_variance}
\Delta N_D^2=\sum_{mn} \Gamma_{mn}=N_D(1+h N_D)    
\end{equation}
is the variance of the total number of detected photons.

\subsection{Single parameter estimation}
If all parameters except separation $d$ are known, the sensitivity matrix become a number
\begin{equation}
    \label{eqn:Md}
    M_d=N_D \sum_m \frac{1}{\varepsilon _m} \left (\frac{\partial \varepsilon_m}{\partial d} \right )^2 + \frac{1}{\Delta N_D^2} \left( \frac{\partial N_D}{\partial d} \right )^2.
\end{equation}
This expression  has two terms, the first term equals $N_D M_\varepsilon$, where
\begin{equation}
\label{eqn:Mepsilon}
    M_\varepsilon = \sum_m \frac{1}{\varepsilon _m} \left(\frac{\partial \varepsilon_m}{\partial d}  \right)^2
\end{equation}
does not depend on the state $\hat \rho_0$ but strongly depends on the measurement basis $\{f_m(\vec r) \}$. 
$M_\varepsilon$ can be called the sensitivity per detected photon of relative intensity measurements (see two-parameters case for the details). In the small photon number limit it coincides with the FI of relative intensity measurement.

The second term
\begin{equation}
\label{eqn:MND}
    M_D=\frac{1}{\Delta N_D^2}  \left(\frac{\partial N_D}{\partial d}  \right)^2,
\end{equation}
does not depend on the individual signals $N_m$, but only on the total one $N_D$, thus it stays the same for any measurement basis $\{f_m(\vec r) \}$ as long as all photons in the image plane are detected. This additional sensitivity $M_D$ occurs due to the interference of mutually coherent sources and the subsequent dependence of the total number of registered photons $N_D$ on the separation $d$. 
The expression  \eqref{eqn:MND} has a self-consistent structure representing the simple error-propagation formula. Variance of the total number of detected photons \eqref{eqn:N_D_variance} depends on the quantum statistics of the source and growth with source bunching.
In the next subsection we show, that this sensitivity of total photon number measurement $M_D$ vanishes, if sources' brightness $N_S$ is unknown.

Total sensitivity $M_d=N_D M_\varepsilon + M_D$ sets lower bound for separation estimator variance 
\begin{equation}
\label{eqn:bound1}
\Delta d^2 \ge \frac{1}{\mu} (M_d)^{-1}=\frac{1}{\mu} \frac{1}{N_D M_\varepsilon + M_D}.
\end{equation}

\subsection{Two parameters estimation}
For better understanding of the physical meaning of values $M_\varepsilon$ and $M_D$ let's consider two-parameter problem, where both separation $d$ and emitted number of photons $N_S$ are unknown. Sensitivity matrix \eqref{eqn:Mmatrix} of estimating a set of variables $q_\alpha=\{d,N_S\}$ reads
\begin{equation} 
     \label{eqn:Mmatrix2}
     M_{\alpha \beta}=
      \begin{pmatrix}
    N_D M_\varepsilon+M_D & ~~\dfrac{1}{\Delta N^2_D} \dfrac{\partial N_D}{\partial N_S} \dfrac{\partial N_D}{\partial d} \\[12pt]
     \dfrac{1}{\Delta N^2_D} \dfrac{\partial N_D}{\partial N_S}\dfrac{\partial N_D}{\partial d} & ~~\dfrac{1}{\Delta N^2_D} \left(\dfrac{\partial N_D}{\partial N_S} \right)^2
        \end{pmatrix},
\end{equation}
where $\partial \varepsilon_m/\partial N_S=0$ is taken into account. In this case variance of $d$ estimation is bounded by 
\begin{equation}
\label{eqn:bound2}
\Delta d^2 \ge \frac{1}{\mu} (M^{-1})_{11}=\frac{1}{\mu} \frac{1}{N_D M_\varepsilon}.
\end{equation}
This bound shows, that without knowing the number of emitted photons $N_S$ one can benefit only from measurement of relative intensities $\varepsilon_m$.
Comparing it to the bound \eqref{eqn:bound1} one can see that knowing the number of emitted photons $N_S$ increases sensitivity of separation estimation by total photon number sensitivity $M_D$. Thus, ignorance of sources brightness  $N_S$  wipes out any possible profit from non-classical statistics of the sources, that are only present in term $M_D$. 

If one uses bucket detection, that corresponds to the detection of all the photons in the image plane, and can be described in this particular case as a single  mode measurement in the principle mode, then the only relative intensity $\epsilon_0=1$ does not depend on parameter, thus $M_\epsilon=0$. One can not estimate separation from this measurement without knowing the number of emitted photons $N_S$. If $N_S$ is known then, as expected, full sensitivity of bucket detection is provided by total photon number detection $M_d=M_D$. 

Note, even though the structure of the sensitivity expression \eqref{eqn:Mmatrix} looks quite intuitive, it was derived specifically for the single mode (fully coherent) case and does not necessarily holds for other cases. For example, the sensitivity of incoherent thermal sources' separation estimation non-linearly depends on the sources brightness \cite{sorelli2021PRA}, although the total number of detected photons does not depend on the separation in this case.

It is worth mentioning, that the used property \eqref{eqn:property} is valid for a quite general class of parameter-estimation schemes, where the parameters are encoded in an arbitrary number of mutually coherent modes, that are subjected to correlated parameter-dependent linear losses. Therefore formula \eqref{eqn:Mmatrix} for the sensitivity matrix of photon counting is valid for this wider class of systems, since the explicit form of coefficients $A_m$ \eqref{eqn:Am} was never used yet. Thus the developed approach can be used for other problems like coherent imaging or quantum sensing in a continuous variable entangled network in the single-mode regime \cite{guo2020distributed}.

\section{Sensitivity of relative intensity measurement $M_\varepsilon$}
Now let us consider separately the two parts of the separation estimation sensitivity \eqref{eqn:Md}. The sensitivity of the relative intensity measurement $M_\varepsilon$ does not depend on the quantum state $\hat \rho_0$, but strongly depends on the measurement basis $\{f_m(\vec r) \}$. As we already mentioned, it equals zero for bucket detection in the principal mode. Here, we consider two more measurement bases.
\subsection{Direct imaging}
For DI the intensity distribution in the image plane reads
\begin{multline}
    I(\vec r)=\kappa N_S \big(u_0^2(\vec r-\vec r_1) \cos^2 \theta+ u_0^2(\vec r-\vec r_2) \sin^2\theta+ \\ + u_0(\vec r-\vec r_1)u_0(\vec r-\vec r_2) \sin 2\theta \cos \phi \big).
\end{multline}
Then the sensitivity of a relative intensity measurement in the continuous limit can be calculated as
\begin{equation}
    \label{eqn:MDI}
    M_\varepsilon^{DI} = \int \frac{1}{i(\vec r)} \left( \frac {\partial i(\vec r)}{\partial d}\right)^2 d \vec r, 
\end{equation}
where $i(\vec r)=I(\vec r)/N_D$, in which we introduce the total detected photon number 
\begin{equation}
    \label{eqn:ND}
    N_D=\int I(\vec r)d \vec r=\kappa N_S (1 + \chi \delta). 
\end{equation}
In turn, we introduce the parameters
\begin{equation}
\label{eqn:chi}
    \chi=\sin 2\theta \cos \phi,
\end{equation}
 and $\delta$ the overlap between the images of the sources
 \begin{equation}
     \delta=\int u_0(\vec r-\vec r_1)u_0(\vec r-\vec r_2) d \vec r.
 \end{equation}
 Expression \eqref{eqn:ND} is valid for any measurement basis in which one can decompose the image mode. Hereinafter we stick to a soft aperture model with a Gaussian PSF
\begin{equation}
    u_0(\vec r)=\sqrt{\frac{1}{2 \pi \sigma^2}} \exp \left [{-\frac{|\vec r|^2}{4 \sigma^2}} \right],
\end{equation}
where $\sigma$ is the width of the PSF. This results in an overlap 
\begin{equation}
    \label{eqn:delta}
\delta= \exp \left[ - \frac{d^2}{8\sigma^2}\right].
\end{equation}

The sensitivity \eqref{eqn:MDI} can be calculated analytically in the cases of in-phase ($\phi=0$) and anti-phase ($\phi=\pi$) sources 
\begin{equation}
\label{eqn:MDI_0pi}
    M_\varepsilon^{DI}\Bigg|_{\phi=0,\pi}= \frac{1}{4\sigma^2} \left(1- \chi \delta + \frac{d^2}{4\sigma^2} \frac{\chi \delta}{1+\chi \delta} \right),
\end{equation}
and in the case of fully asymmetric sources ($\theta=0$, which is equivalent to a single source's centroid estimation) giving the well known result
\begin{equation}
    \label{eqn:MDI_0theta}
    M_\varepsilon^{DI}\Bigg|_{\theta=0}=\frac{1}{4\sigma^2}.
\end{equation}
For this case $\chi=0$, meaning that eqn. \eqref{eqn:MDI_0pi} is also valid for $\theta=0$. For other values of $\phi$ and $\theta$ the sensitivity $M_\varepsilon^{DI}$ is calculated numerically.

\begin{figure}[ht]
\begin{tabular}{c}
  \includegraphics[width=0.65 \linewidth]{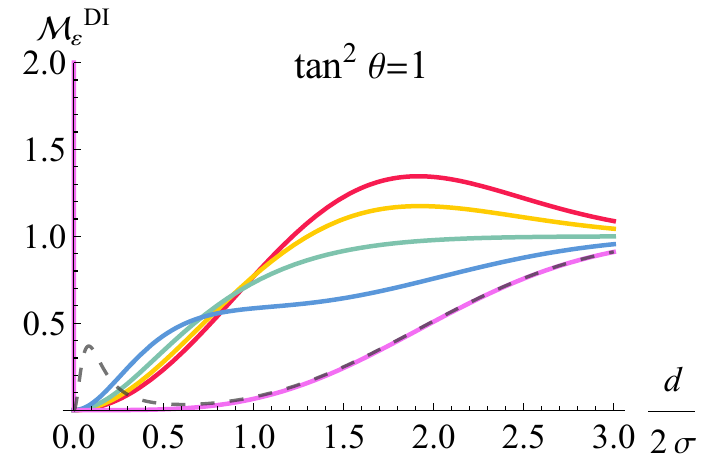} \\
   \includegraphics[width=0.65 \linewidth]{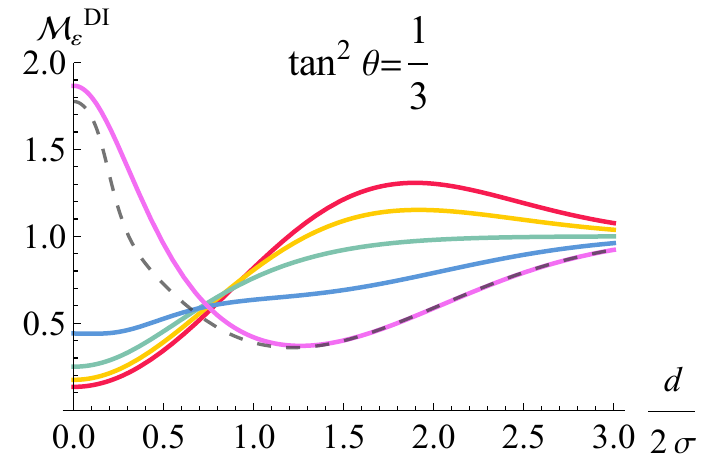}
\end{tabular}
 \begin{tabular}{c}
      \includegraphics[width=0.292 \linewidth]{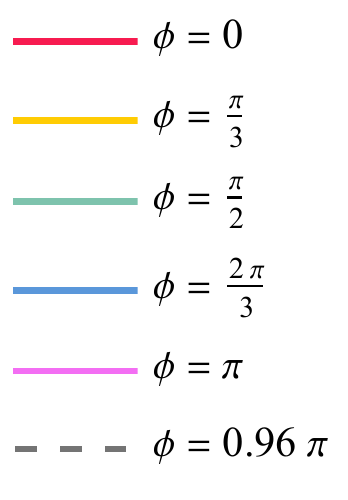}
 \end{tabular}
 \caption{\label{fig:plotDI} Normalised sensitivity per emitted photon of relative intensity distribution direct measurement   $\mathcal{M}_\varepsilon^{DI}$. Top panel corresponds to the sources with equal intensity and parameter $\chi=\{1, 1/2, 0,-1/2, -1, -0.99\}$, bottom panel -- to asymmetric case and  $\chi=\{0.87, 0.43, 0, -0.43, -0.87, -0.86\}$.}
 \end{figure}

The variable $M_\epsilon$ corresponds to the sensitivity per detected photon. The sensitivity per emitted photon is instead given by $N_D M_\epsilon / N_S$. We additionally normalise this value over transmissivity $\kappa$ and multiply by $4\sigma^2$ to remove dependence on these parameters. A plot of the resulting normalised sensitivity 
\begin{equation}
    \label{eqn:Mepsilon_N}
    \mathcal{M}_\varepsilon= \frac{4\sigma^2}{\kappa}\frac{N_D M_\varepsilon}{N_S}
\end{equation}
for the case of DI is presented on fig. \ref{fig:plotDI}. 

On the top inset of fig.~\ref{fig:plotDI} one can see that direct measurement of the relative intensity distribution in the case of equally bright sources leads to low sensitivity for small separation and, as expected, to ``Raleigh's curse'', i.e., vanishingly small sensitivity for infinitely small $d$. In case of asymmetric sources (bottom of fig.~\ref{fig:plotDI}) Raleigh's curse does not occur for direct imaging, although the sensitivity is quite small for small separation, unless the sources are nearly in anti-phase.
 
\subsection{Spatial demultiplexing}
Another imaging technique under study is SPADE. It was shown that the sensitivity of photon counting in HG modes saturates QFI for the estimation of two equally-bright incoherent thermal sources' separation in case of Gaussian PSF \cite{sorelli2021PRA}. This measurement also beats DI in the asymmetric case (unequally bright sources), though the optimality of the HG basis was never proved for this case.

Here, we analyse the sensitivity of photon counting in HG modes in the case of fully coherent bright sources in an arbitrary quantum state. We assume to know all the sources' parameters except the separation (in \eqref{eqn:Md} and \eqref{eqn:Mmatrix2}), and the total brightness $N_S$ (only in \eqref{eqn:Mmatrix2}). This implies that the position of the centroid $(\vec r_1 + \vec r_2)/2$ 
and the orientation of the pair of sources are assumed to be known prior to the measurement. Often these parameters, if unknown, can be estimated with an additional preparatory DI measurement. However, one should keep in mind that such an estimation has a finite precision, and, for an asymmetric source, it is correlated with the estimation of the separation itself. Under these assumptions, the sources' positions equal to $\vec r_{1,2}=\{\pm d/2,0 \}$.  And the measurement basis can be chosen aligned with the image centroid, i.e. the measurement HG mode basis reads
\begin{equation}
f_m(x,y)= \frac{1}{\sqrt{2^m m!}} H_m \left( \frac{x}{\sqrt{2} \sigma} \right) u_0 \left(\sqrt{x^2+y^2}\right), 
\end{equation}
where $H_m$ are Hermite polynomials. Calculating the overlaps with the image modes we find that the coefficients $A_m$ \eqref{eqn:Am} are given by
\begin{equation}
    \label{eqn:Am_SPADE}
    A^{HG}_m= \sqrt \kappa  \Big((-1)^m \cos \theta + e^{i \phi} \sin \theta \Big) ~\beta_m \left( \frac{d}{4 \sigma} \right),
\end{equation}
where 
\begin{equation}
    \label{eqn:beta_HG}
    \beta_m (x_0)=
    e^{-\frac{x_0^2}{2}} \frac{x_0^m}{\sqrt{m!}}.
\end{equation}
This allows to find the mean photon numbers in the measurement modes \eqref{eqn:Nm}
\begin{equation}
    N^{HG}_m=N_S\Big(1+(-1)^m \chi\Big) ~\beta_m^2 \left( \frac{d}{4\sigma} \right).
\end{equation}
These can be normalised with respect to total number of detected photons $N_D$ \eqref{eqn:ND} to calculate the sensitivity \eqref{eqn:Mepsilon} of separation estimation from  measured relative photon numbers $\varepsilon_m$. In case of infinitely many HG modes being measured this sensitivity reads
\begin{equation}
\label{eqn:MHG}
    M_\varepsilon^{HG}=\frac{1}{4\sigma^2} \left(1- \chi \delta + \frac{d^2}{4\sigma^2} \frac{\chi \delta}{1+\chi \delta} \right).
\end{equation}
Note, that $M_\varepsilon^{HG}$ depends only on the combination $\chi$ of the parameters $\theta$ and $\phi$. 

One can notice that the expression for SPADE's sensitivity $M_\varepsilon^{HG}$ \eqref{eqn:MHG} coincides with the sensitivity of DI $M_\varepsilon^{DI}$ in cases of in-phase, anti-phase \eqref{eqn:MDI_0pi}, and fully asymmetric sources \eqref{eqn:MDI_0theta}. It is possible to show, that in the general case 
\begin{equation}
\label{eqn:theorem}
M_\varepsilon^{HG} \ge M_\varepsilon^{DI},    
\end{equation}
and the inequality is saturated only if $\phi=0$ or $\pi$ or $\theta=0$ (see Appendix \ref{app:HGvsDI} for the proof). Meaning that measurements in the HG basis are always more (or equally) sensitive then DI for separation estimation. However,in the asymmetric case ($\theta \ne \pi/4$), we cannot prove the optimality of this measurement, since the QFI is unknown.

\begin{figure}[ht]
  \includegraphics[width=0.65 \linewidth]{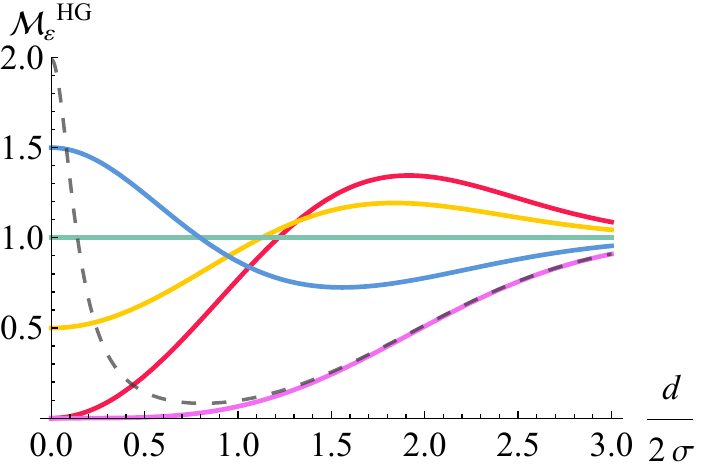}    \includegraphics[width=0.292 \linewidth]{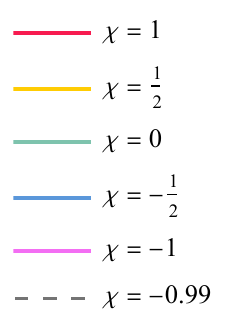}
 \caption{\label{fig:plotHG} Normalised sensitivity per emitted photon of relative intensity measurement in HG modes $\mathcal{M}_\varepsilon^{HG}$.}
 \end{figure}

In fig. \ref{fig:plotHG}, we plot the normalised sensitivity \eqref{eqn:Mepsilon_N} of the relative intensity measurement in the HG modes. One can see that Raleigh curse is still present for the symmetric in-phase ($\chi=1$) and anti-phase ($\chi=-1$) cases, which is not surprising, since SPADE for this cases is as sensitive as DI. Note, however, that even a small deviation from the symmetric anti-phase case ($\chi=-1$) leads to a significant sensitivity increase for small separations (see. dashed line on fig. \eqref{fig:plotHG}).

\begin{figure*}[ht]
  \includegraphics[width=0.28 \linewidth]{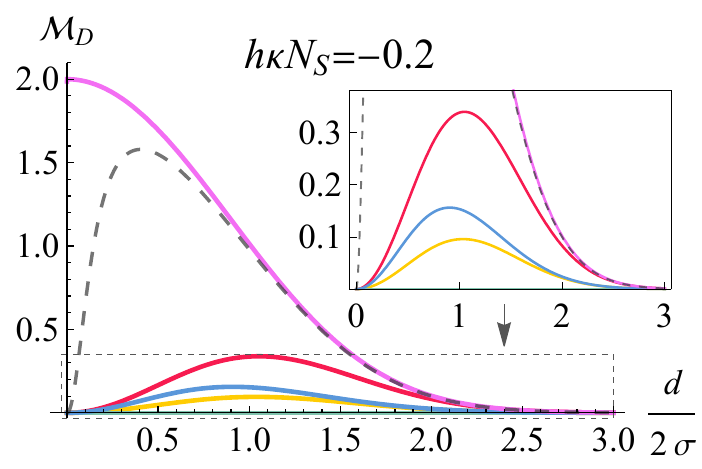} 
  \includegraphics[width=0.28 \linewidth]{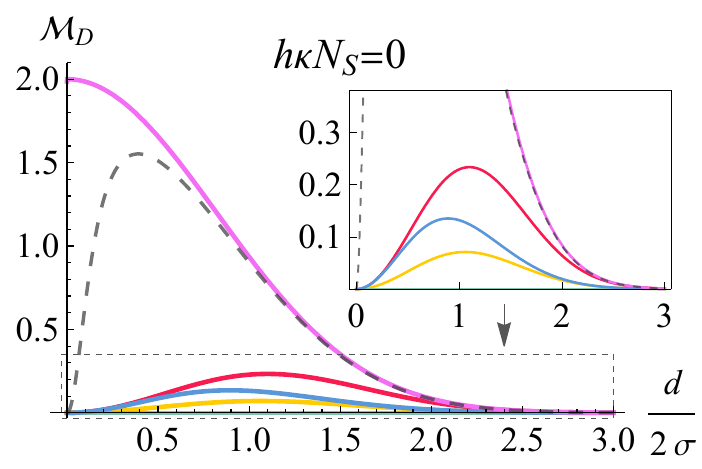} 
  \includegraphics[width=0.28 \linewidth]{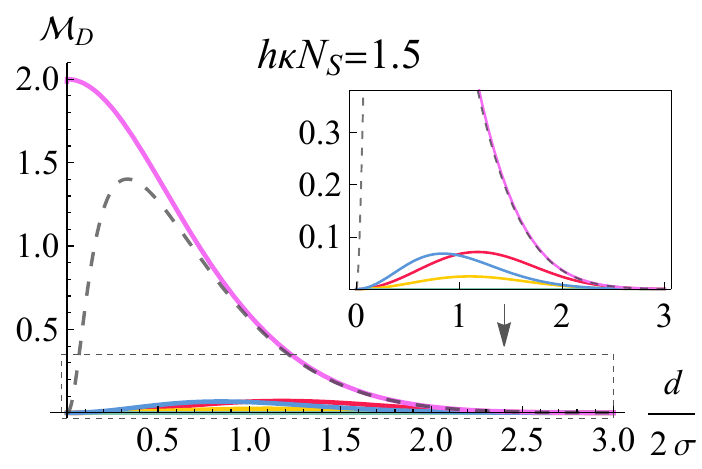} 
  ~~\includegraphics[width=0.13 \linewidth]{pics/legHG.pdf}
  \caption{\label{fig:plotMD} Normalised sensitivity of total photon number detection $\mathcal{M}_D$. From left to right: Fock ($\kappa=0.2$), coherent and thermal state ($N_S=1.5/\kappa$) of the mode $\hat s_0$.}
 \end{figure*}

The sensitivity in the case $\chi=0$ (that can correspond to the mutual phase $\phi=\pi/2$) does not depend on the separation $d$ and coincide with sensitivity and QFI in case of weak uncorrelated thermal sources \cite{pirandola,sorelli2021PRL}. However for incoherent thermal sources the QFI per emitted photon drops with growing number of photons, when $\mathcal{M}_\varepsilon^{HG}$ for correlated sources does not depend on $N_S$ for any photon statistics of the source.

\section{Sensitivity of total intensity measurement $M_D$}
Having an expression for the total number of detected photons \eqref{eqn:ND} one can analytically calculate the total photon-number sensitivity \eqref{eqn:MND}
\begin{equation}
\label{eqn:MD_explicit}
    M_D=\frac{\kappa N_S}{4 \sigma^2} \frac{\delta^2 \chi^2}{(1+\delta \chi)+h \kappa N_S(1+\delta \chi)^2} \left(\frac{d}{2\sigma} \right)^2.
\end{equation}

This expression is valid for any detection basis in which the image mode can be decomposed, i.e. for both DI and SPADE. It also includes all quantum states of the sources (as long as the sources are mutually coherent) via the coefficient  $h=g^{(2)}-1=(\Delta N_S^2 - N_S)/N_S^2$. From \eqref{eqn:MD_explicit} it is obvious, that anti-bunched states of $\hat s_0$ ($h<0$), leading to entanglement in modes $\hat s_{1,2}$, provide a better sensitivity then bunched states ($h>0$) of $\hat s_0$, which corresponds to classical correlations in $\hat s_{1,2}$. This is a natural result, since a lower photon number variance in $\hat s_0$ leads to a smaller variance of $N_D$ and hence a higher sensitivity of the $N_D$ measurement. 

Here we consider the sensitivity of separation estimation from a measured total intensity $N_D$ for different quantum statistics of the sources. We are interested in the normalised sensitivity per emitted photon 
\begin{equation}
\label{eqn:MD_N}
    \mathcal{M}_D=\frac{4\sigma^2}{\kappa}\frac{M_D}{N_S}.
\end{equation}
The characteristics of the source statistics only appear in the combination $h \kappa N_S$, furthermore $\mathcal{M}_D$ also depend on $\chi$ and the separation $d$.

We explore the impact of the source statistics by studying various common initial states.
\paragraph{Fock state.}
We start by considering the most sensitive case, when the mode $\hat s_0$ is maximally anti-bunched, i.e. the Fock state. In this case $h=-1/N_S$, and the combination $h \kappa N_S=-\kappa$. On the left panel of fig. \ref{fig:plotMD}, we plot the sensitivity $\mathcal{M}_D$ \eqref{eqn:MD_N} with $\kappa=0.2$  (used model of linear losses requires $\kappa \ll 1$). Note that, for the Fock state, the sensitivity per emitted photon does not depend on the number of photons $N_S$.

\paragraph{Poisson statistics.}
If the mode $\hat s_0$ is in the coherent state, i.e. has Poisson statistics, then the state of the modes $\hat s_{1,2}$ is a direct product of coherent states. For this case the parameter $h=0$, and the sensitivity per emitted photon also does not depend on the source intensity. The normalised sensitivity for this case is plotted on the middle inset of fig. \ref{fig:plotMD}.

\paragraph{Thermal statistics.}
Finally we consider a thermal state of the mode $\hat s_0$, that leads to correlated thermal states in modes $\hat s_{1,2}$. For thermal statistics $h=1$. In the small photon number limit ($N_S \to 0$) the sensitivity per emitted photon $\mathcal{M}_D$ coincides with the coherent case, and for high photon number ($N_S \to \infty$) it vanishes ($\mathcal{M}_D \to 0$). We plot the normalised total photon-number sensitivity for correlated thermal sources for $\kappa N_S=1.5$ on the right inset of the fig. \ref{fig:plotMD}. 

For all the considered cases the total photon number sensitivity is high in the case of small separation between symmetric anti-phase sources ($\chi=-1$). This occurs due to destructive interference of mutually coherent anti-phase sources, which leads to zero intensity in the image plane if equally bright sources coincide, and non-zero total intensity in presence of finite separation between the sources. For any other case sensitivity $M_D$ vanishes for zero separation. In the case of $\chi=0$ the total photon number $N_D$ \eqref{eqn:ND} does not depend on the parameter and $M_D=0$.

 \begin{figure*}[ht]
  \includegraphics[width=0.28 \linewidth]{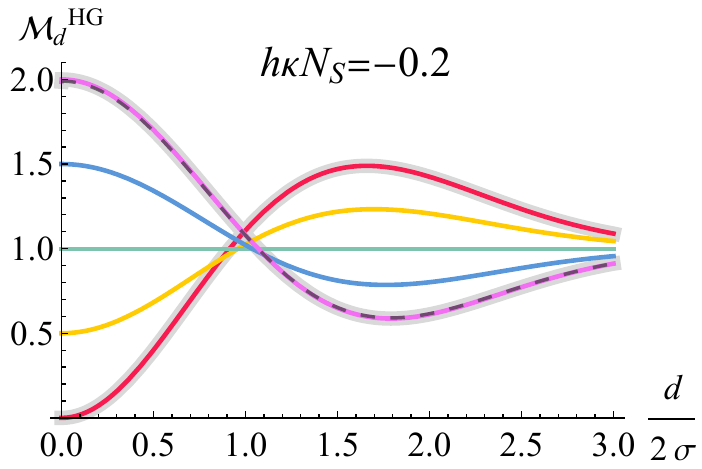} 
  \includegraphics[width=0.28 \linewidth]{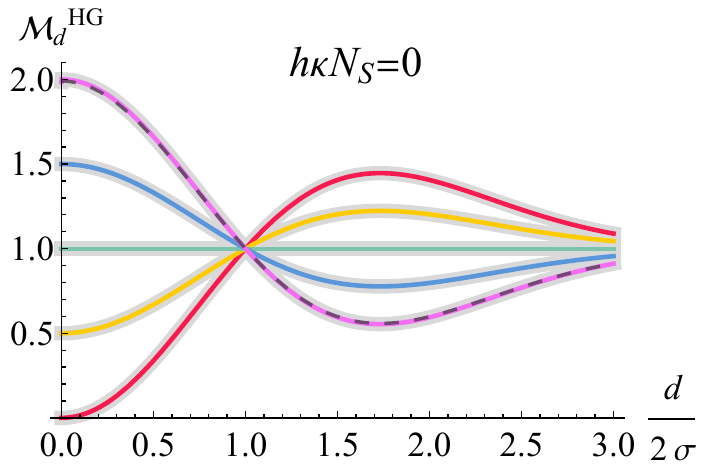} 
  \includegraphics[width=0.28 \linewidth]{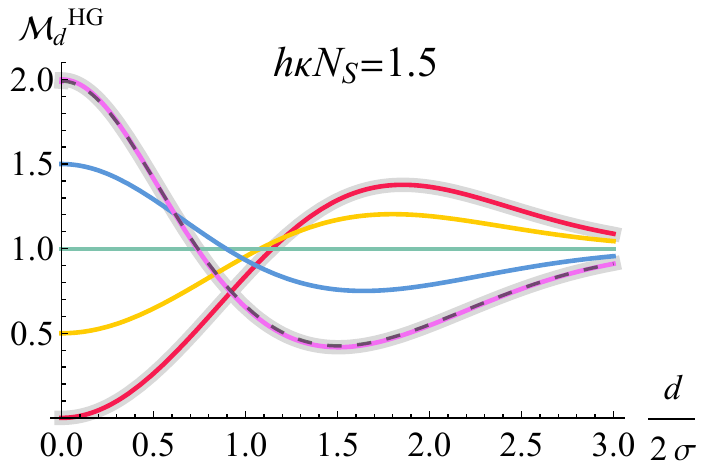} 
  ~~\includegraphics[width=0.13 \linewidth]{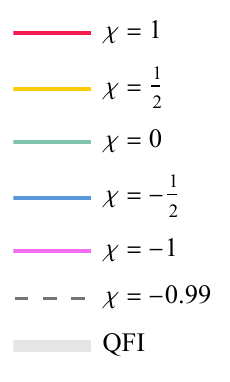}
  \caption{\label{fig:plotMd_Final} Normalised total separation estimation sensitivity $\mathcal{M}^{HG}_d$ via SPADE in HG basis. From left to right: Fock ($\kappa=0.2$), coherent and thermal state ($N_S=1.5/\kappa$) of the mode $\hat s_0$. Cases with known QFI are highlighted with gray, QFI and $M_d^{HG}$ coincide for all of them.}
 \end{figure*}

\section{Comparison with Quantum Fisher information}
Here we analyze the total separation estimation sensitivity $M_d=N_D M_\varepsilon + M_D$ and it's normalized version
\begin{equation}
\label{eqn:Md_N}
     \mathcal{M}_d= \frac{4\sigma^2}{\kappa}\frac{ M_d}{N_S}=\mathcal{M}_\varepsilon+\mathcal{M}_D.
\end{equation}
We compare the sensitivity of SPADE with the ultimate limit set by the QFI in those cases where the QFI is explicitly known.

\paragraph{Fock state.} The first example we consider is a Fock state of the mode $\hat s_0$. Splitting it on the beamsplitter result in entanglement of modes $\hat s_{1,2}$. Particularly for the symmetric beamsplitter and mutual phase $\phi=0,\pi$ state of the sources takes form
\begin{equation}
\label{eqn:entangled_states}
    | \psi \rangle _{s_1 s_2} = \frac{1}{\sqrt{2^{N_S}}} \sum_{j=0}^{N_S} {(\pm 1)^{N_S-j}} | j \rangle_{s_1} | N_S-j \rangle_{s_2}.
\end{equation}
One of these states corresponds to the maximal QFI of separation estimation (which one depends on the transmissivity $\kappa$ and separation $d$)  \cite{pirandola}. The analytical expression obtained for the SPADE sensitivity $M_d^{HG}$ coincides with the QFI for the states \eqref{eqn:entangled_states}. For other values of mutual phase $\phi$, or asymmetrically split Fock states, the QFI is not calculated explicitly to the best of our knowledge.  
Plots of the sensitivity of SPADE and the QFI for split Fock states are presented on the left inset of fig. \ref{fig:plotMd_Final}.

\begin{figure}[h]
  \includegraphics[width=0.65 \linewidth]{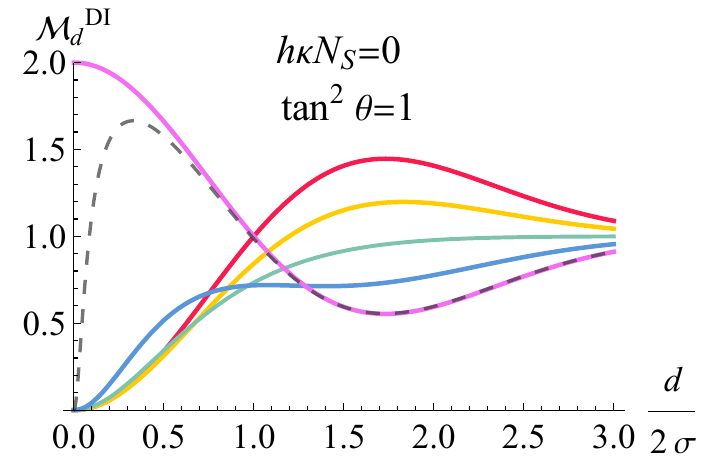} 
  ~~\includegraphics[width=0.27 \linewidth]{pics/legHG.pdf}
  \caption{\label{fig:plotMd_DI_Final} Total separation estimation sensitivity of DI $M_d^{DI}$ for equally bright sources with Poisson statistics.}
 \end{figure}

\paragraph{Poisson statistics.} Since for the Poisson photon number statistics different detection events are independent from each other, this case can be considered in the small photon number limit without loosing any generality and then generalised to higher photon numbers. The QFI for an arbitrary mutual coherence $\gamma$ was explicitly calculated in the ref. \cite{kurdzialek2021sources} in the single-photon subspace, and its analytical expression for any $\gamma=e^{i \phi}$ fully coincides with the sensitivity $M_d$ calculated for Poisson statistics ($h=0$). The dependence of normalised QFI and $\mathcal{M}_d^{HG}$ on the separation are presented in the middle inset of fig. \ref{fig:plotMd_Final}, coinciding with each other.  We also plot sensitivity of DI for symmetric sources with Poisson statistics on the fig. \ref{fig:plotMd_DI_Final}

\paragraph{Thermal statistics.} Another example we consider is that of correlated thermal sources. The QFI for arbitrarily bright correlated thermal sources that are in-phase or anti-phase is calculated in ref. \cite{pirandola}. It coincides with the sensitivity $M_d^{HG}$ introduced here. With increasing intensity of the correlated thermal source, the sensitivity per photon $\mathcal{M}_d$ drops tending to $\mathcal{M}_\varepsilon$ (fig. \ref{fig:plotHG}).

One can notice, that distinctive behaviour of $\mathcal{M}^{HG}_\epsilon$ and $\mathcal{M}^{HG}_D$ for $\chi=0.99$ cancel each other after summation for any source statistics, and the resulting sensitivity of SPADE $\mathcal{M}^{HG}_d$ is continuous over change of $\chi$, which is not the case for DI. 

Fig. \ref{fig:plotMd_Final} also shows, that source statistics does not influence the sensitivity dramatically if the sources are mutually coherent. At the same time comparing fig. \ref{fig:plotMd_Final} and fig. \ref{fig:plotMd_DI_Final} one can see that choice of the measurement basis, i.e. HG modes (SPADE) or pixel modes (DI), result in sufficient difference in sensitivity, unless sources are in-phase or anti-phase.

\section{Conclusion}
We presented a general approach to analyse parameter estimation problems based on photon-counting in mutually coherent modes. The considered sensitivity measure based on the method of moments showed to be a very efficient and practical tool for analyzing this class of problems. 

Specifically we have considered in details the problem of separation estimation of two mutually coherent sources. By calculating the moment-based sensitivity, we analytically showed an advantage of the spatial mode demultiplexing measurement over the direct imaging for the considered class of states. We showed that sensitivity of spatial mode demultiplexing saturates the QFI for the cases where it is known, including examples of non-Gaussian entangled states.

Obtained sensitivity expression consists of two terms, that are corresponding to the measurement of the relative photon numbers and the total number of photons. The first one strongly depends on the measurement basis, the second one on quantum statistics of the sources. Moreover, the second term vanishes in case of unknown brightness of the sources, wiping out any profit from the sources anti-bunching. The sensitivity from the total photon number measurement is also negligible for intense bunched states, due to the high noise in the total number of photons. 

The moment-based sensitivity approach and results obtained in this research can be applied to other parameter estimation problems with mutually coherent or single-mode sources, like coherent imaging or distributed quantum sensing with mutually coherent probes.

 \begin{acknowledgments} 
 I.K. acknowledges the support from the PAUSE National programme.  M.G. acknowledges funding by the LabEx ENS-ICFP:ANR-10-LABX-0010/ANR-10- IDEX-0001-02 PSL*. This work was partially funded by French ANR under COSMIC project (ANR-19-ASTR- 0020-01). This work received funding from the European Union’s Horizon 2020 research and innovation programme under Grant Agreement No. 899587. This work was supported by the European Union’s Horizon 2020 research and innovation programme under the QuantERA programme through the project ApresSF.
 \end{acknowledgments}

\bibliography{references}

\appendix 
\section{Proof of SPADE advantage over DI}
\label{app:HGvsDI}
\textbf{Theorem.} 
\begin{equation}
\label{eqn:deltaM}
    \Delta M = M_d^{HG}-M_d^{DI} \ge 0,
\end{equation}
i.e. separation estimation sensitivity with SPADE measurements in the HG basis outperforms DI for any pair of mutually coherent sources. The inequality \eqref{eqn:deltaM} is saturated only in cases $\phi=0$, $\phi=\pi$ or $\theta=0$.

\textit{Proof.} To estimate the difference $\eqref{eqn:deltaM}$ one can rewrite the sensitivity \eqref{eqn:Md} in the following way 
\begin{equation}
M_d=\sum_m \frac{1}{N _m} \left (\frac{\partial N_m}{\partial d} \right )^2 - \frac{h}{1+h N_D\\
} \left( \frac{\partial N_D}{\partial d} \right )^2,
\end{equation}
where the second term is independent of the measurement basis. Then 
\begin{equation}
\Delta M = M_0^{HG}(\theta,\phi)-M_0^{DI}(\theta,\phi),
\end{equation}
where
\begin{equation}
\label{eqn:M0Def}
M_0 (\theta, \phi)=\sum_m \frac{1}{N _m} \left (\frac{\partial N_m}{\partial d} \right )^2.
\end{equation}

From the equality $M_\varepsilon^{DI}\big|_{\phi=0}=M_\varepsilon^{HG}\big|_{\phi=0}$ (see eqn. \eqref{eqn:MDI_0pi} and \eqref{eqn:MHG}) follows that 
\begin{equation}
M_0^{DI} (\theta, 0) = M_0^{HG} (\theta, 0),
\end{equation}
since the difference between $M_\varepsilon$ and $M_0$ is basis independent. At the same time $M_0^{HG}$ depends only on the parameter combination $\chi=\sin 2\theta \cos \phi$. Then, for any $\phi \le \pi/2$, one can use the following chain of equalities:
\begin{equation}
M_0^{HG} (\theta, \phi) = M_0^{HG} (\theta_1,0) = M_0^{DI} (\theta_1,0), 
\end{equation}
where $\sin 2\theta_1=\sin 2 \theta \cos \phi$. One can redo all the following analysis for $\phi \ge \pi/2$ using the fact that $M_\varepsilon^{DI}\big|_{\phi=\pi}=M_\varepsilon^{HG}\big|_{\phi=\pi}$, therefore results are true for any value of $\phi$.

Thus the difference in sensitivity can be expressed as
\begin{equation}
\label{eqn:deltaM2}
    \Delta M = M_0^{DI} (\theta_1,0)- M_0^{DI} (\theta,\phi).
\end{equation}

For continuous DI the formula \eqref{eqn:M0Def} takes the form
\begin{equation}
    M_0^{DI} (\theta,\phi)=\int \frac{1}{I_{\theta,\phi}(\vec r)} \left( \frac {\partial I_{\theta,\phi}(\vec r)}{\partial d}\right)^2 d \vec r
\end{equation}
where $I_{\theta,\phi}(\vec r)=|E_{\theta,\phi}(\vec r)|^2$ and 
\begin{equation}
\label{eqn:E}
    E_{\theta,\phi}(\vec r)=u_0(\vec r - \vec r_1) \cos \theta +u_0(\vec r - \vec r_2) e^{i \phi} \sin \theta. 
\end{equation}
A simple transformations leads to
\begin{multline}
\label{eqn:M0DI}
    M_0^{DI} (\theta, \phi) = 2 \operatorname{Re} \Bigg[ \int  \Bigg( \frac{E_{\theta,\phi}^*(\vec r)}{E_{\theta,\phi}(\vec r)} (E'_{\theta,\phi}(\vec r))^2 + \\
    +  \left|E'_{\theta,\phi}(\vec r)\right|^2  \Bigg) d\vec r \Bigg],
\end{multline}
where $E'_{\theta,\phi}(\vec r)$ stands for derivative with respect to separation $d$. Using inequality
\begin{equation}
\label{eqn:integral_ineq}
    \operatorname{Re} \left[\int g(x)  dx\right] \le \int |g(x)| dx,
\end{equation}
for the first term in \eqref{eqn:M0DI} we find that
\begin{equation}
\label{eqn:M0DIless}
     M_0^{DI} (\theta, \phi) \le 4 \int \left|E'_{\theta,\phi}(\vec r)\right|^2  d\vec r.
\end{equation}

Since $E_{\theta_1,0}(\vec r) \in \mathbb{R}$, then \eqref{eqn:M0DI} in this special case simplifies to
\begin{equation}
\label{eqn:M0HG}
     M_0^{DI} (\theta_1, 0) = 4 \int \left|E'_{\theta_1,0}(\vec r)\right|^2  d\vec r.
\end{equation}
By using the explicit expression for the electric field \eqref{eqn:E}, we can calculate the integral from the right part of \eqref{eqn:M0DIless}. The result depends only on $\chi=\sin 2 \theta \cos \phi$. This means that the right parts of \eqref{eqn:M0DIless} and \eqref{eqn:M0HG} are equal, resulting in
\begin{equation}
    \frac{\Delta M}{4} \ge  \int \left|E'_{\theta_1,0}(\vec r)\right|^2  d\vec r -  \int \left|E'_{\theta,\phi}(\vec r)\right|^2  d\vec r=0.
\end{equation}

The inequality \eqref{eqn:integral_ineq} is only saturated if $g(x)=|g(x)|$, i.e.
\begin{equation}
\label{eqn:condition}
    \frac{E_{\theta,\phi}^*(\vec r)}{E_{\theta,\phi}(\vec r)} (E'_{\theta,\phi}(\vec r))^2=\left| \frac{E_{\theta,\phi}^*(\vec r)}{E_{\theta,\phi}(\vec r)} (E'_{\theta,\phi}(\vec r))^2 \right|
\end{equation}
for any $\vec r$. 
This equality holds only in cases $\phi=0$, $\phi=\pi$ or $\theta=0$. 

\begin{figure}[ht]
\begin{tabular}{c}
  \includegraphics[width=0.65 \linewidth]{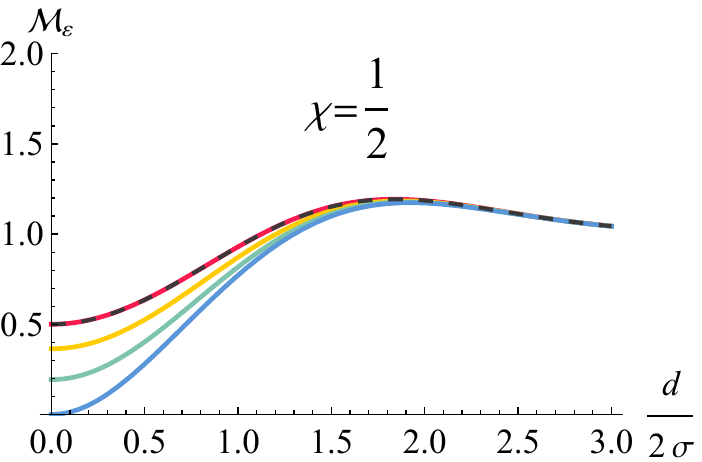} \\
  \includegraphics[width=0.65 \linewidth]{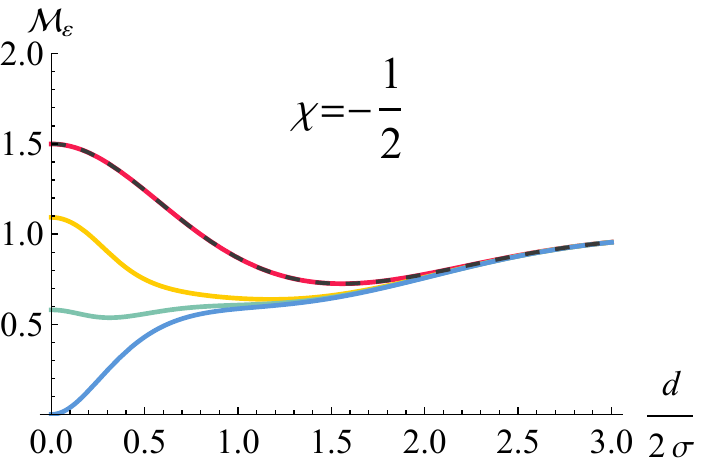}
\end{tabular}
 \begin{tabular}{c}
      \includegraphics[width=0.3 \linewidth]{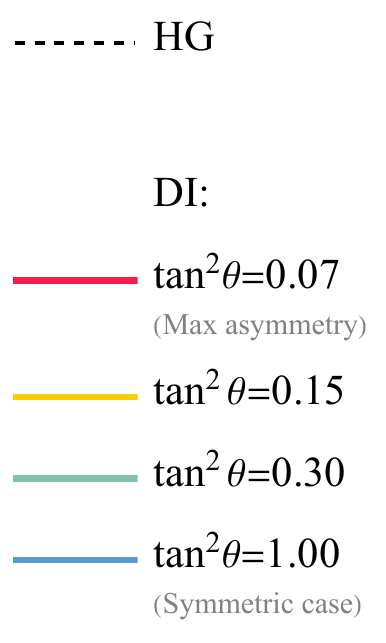}
 \end{tabular}
 \caption{\label{fig:plotHGvsDI} Sensitivity per emitted photon of relative intensity measurement. Comparison DI and SPADE technique.}
 \end{figure}

On the figure \ref{fig:plotHGvsDI} you can find comparison of the DI and SPADE sensitivities, built with fixed combinations $\chi=\sin 2\theta \cos \phi$ but with a different ratio between parameters $\phi$ and $\theta$. These plots illustrate all given relations between $M_d^{HG}$ and $M_D^{DI}$.

\end{document}